\documentclass[letterpaper]{article} 
\usepackage{aaai24}  
\usepackage{times}  
\usepackage{helvet}  
\usepackage{courier}  
\usepackage[hyphens]{url}  
\usepackage{graphicx} 
\usepackage{natbib}  
\usepackage{caption} 
\usepackage{algorithm}
\usepackage{algorithmic}
\usepackage{subcaption}

\usepackage{array}
\usepackage{subfig}
\graphicspath{ {./images/} }
\usepackage{enumitem} 
\newcolumntype{P}[1]{>{\centering\arraybackslash}p{#1}}
 \usepackage{booktabs} 

\usepackage[title]{appendix}

\usepackage{newfloat}
\usepackage{listings}
\DeclareCaptionStyle{ruled}{labelfont=normalfont,labelsep=colon,strut=off} 
\floatstyle{ruled}
\newfloat{listing}{tb}{lst}{}
\floatname{listing}{Listing}
\title{``I don't see myself represented here at all": \\User Experiences of Stable Diffusion Outputs Containing Representational Harms across Gender Identities and Nationalities}
\author{
    Sourojit Ghosh\equalcontrib,
    Nina Lutz\equalcontrib,
    Aylin Caliskan
}
\affiliations{
    University of Washington, Seattle\\
    \{ghosh100, ninalutz, aylin\}@uw.edu
}
\begin{document}

\maketitle

\begin{abstract}
Though research into text-to-image generators (T2Is) such as Stable Diffusion has demonstrated their amplification of societal biases and potentials to cause harm, such research has primarily relied on computational methods instead of seeking information from real users who experience harm, which is a significant knowledge gap. In this paper, we conduct the largest human subjects study of Stable Diffusion, with a combination of crowdsourced data from 133 crowdworkers and 14 semi-structured interviews across diverse countries and genders. Through a mixed-methods approach of intra-set cosine similarity hierarchies (i.e., comparing multiple Stable Diffusion outputs for the same prompt with each other to examine which result is `closest' to the prompt) and qualitative thematic analysis, we first demonstrate a large disconnect between user expectations for Stable Diffusion outputs with those generated, evidenced by a set of Stable Diffusion renditions of `a Person' providing images far away from such expectations. We then extend this finding of general dissatisfaction into highlighting representational harms caused by Stable Diffusion upon our subjects, especially those with traditionally marginalized identities, subjecting them to incorrect and often dehumanizing stereotypes about their identities. We provide recommendations for a harm-aware approach to (re)design future versions of Stable Diffusion and other T2Is. 
\end{abstract}

\section{Introduction}

Since 2022, there has been a meteoric rise in the usage of Text-to-image generators (T2Is), paralleled by research into them. T2Is like Stable Diffusion take textual prompts from users and generate high-resolution images, with little to no barrier of entry for use. T2Is have been well-studied in AAAI/ACM spaces, especially evaluating potential biases and harms that they embed. However, such research \cite[e.g.,][]{bianchi2023easily, ghosh2023person, luccioni2023stable} typically takes algorithmic or researcher-evaluated techniques to labeling biases and resultant harms. Often unheard remains the voice of the \textit{user}, humans who use tools like Stable Diffusion and services that embed such tools, at the receiving end of \textit{representational harms} which perpetuate negative/unwelcome stereotypes \cite{barocas2017problem}. In this paper, we highlight such voices. 

We examine user perspectives on human faces generated by Stable Diffusion, specifically outputs of prompts containing aspects of identities that users themselves hold. Using a combination of qualitative thematic analysis \cite{braun2006using} and normalized cosine similarity comparisons of generated images -- a 0-1 increasing metric comparing image similarity \cite{singhal2001modern} -- across 133 crowdsourced sessions and 14 user interviews engaging with Stable Diffusion outputs around 136 prompts and 50 images per prompt (see Section \ref{subsec:prompt-image}), we make two novel contributions: 

\begin{itemize}
    \item Through the largest-to-date human subjects study of T2Is for academic research, we reveal a large gap between user expectations of Stable Diffusion outputs with those generated. By comparing 133 Prolific users' expectations of Stable Diffusion renditions of `a Person' and 14 interviews of users expecting to see their own identities well-depicted by Stable Diffusion with intra-set cosine similarity (comparing every Stable Diffusion output to others for the same prompt \cite{ghosh2023person}) of outputs for 28 unique prompts (50 images/prompt), we observe how outputs with highest intra-set cosine similarity scores moderately/poorly align with user expectations, whereas outputs highly preferred by users score near the bottom of intra-set cosine similarity rankings. This highlights a usability error within Stable Diffusion, which is concerning given its significant global usage.
    
    \item Having identified a general pattern of user expectations of Stable Diffusion outputs being distant from actual outputs, we demonstrate how users perceive significant representational harms being caused by Stable Diffusion depictions of aspects of their own identities. Through 14 interviews with current users of Stable Diffusion spanning multiple countries, gender identities, cultures, and backgrounds, we document how Stable Diffusion outputs cause the harms of stereotyping, disparagement, dehumanization, and erasure \cite{dev2021measures}. Through qualitative thematic analysis \cite{braun2006using} of interviews, we provide first-hand accounts of users with marginalized and minoritized identities in their local or global contexts, such as people of color in the US or nonbinary/trans people, experiencing the bulk of such harms, as they are dissatisfied and angered by outputs supposed to reflect their identities accurately.     
\end{itemize}

We advocate for proactive community-centered and harm-aware approaches towards Stable Diffusion and, more broadly, T2I design. At a time when design and proliferation of T2Is is ubiquitous by multiple organizations worldwide trying to out-innovate each other, we advocate for T2I design practices foregrounding individuals' agencies and values, rather than retroactively analyzing harms caused.

\section{Related Work}\label{sec:related-work}
In this section, we overview T2Is, and discuss potential representational harms that their outputs can cause. 

\subsection{T2Is, and Stable Diffusion}\label{subsec:rw-stable-diffusion}

Text-to-image generators (T2Is) take in text-based input prompts and return images as outputs. Developments of T2Is have skyrocketed since the advent of diffusion models which go beyond older image dataset and class based supervised paradigms to unsupervised (implicit supervision through language) paradigms by self-generating new data/images based on training data, and then adding in the generated data into training data \cite{dhariwal2021diffusion}. 

We specifically focus on the T2I \textit{Stable Diffusion}, ``a latent text-to-image diffusion model capable of generating photo-realistic images given any text input, [which] empowers billions of people to create stunning art within seconds" \cite{stablediffusion_history}. Its architecture consists of ``a variational autoencoder, forward and reverse diffusion, a noise predictor, and text conditioning," \cite{aws23} outputting 512x512 images in response to text prompts. It is trained on the 2B English subset of the LAION-5B dataset \cite{schuhmann2022laion}, covering 2 billion CLIP-filtered text-image pairs curated from publicly available images. Stable Diffusion, released in 2022 by Stability AI, is available for personal and commercial use to millions of people worldwide \cite{stablediffusionusage}. This popularity and open-source nature has made Stable Diffusion a common subject of T2I research, including downstream applications \cite{lee2023diffusion,liu2023application} and comparison across other T2Is like Dall-E or Midjourney \cite{borji2022generated,rombach2022high}. 

Extensive work has also focused on stereotypes embedded and the societal impact of stereotype propagations \cite[e.g.,][]{ghosh2023person,luccioni2023stable,qadri2023ai}, and development of mitigation strategies such as Safe Latent Diffusion \cite{schramowski2023safe}. While some such studies are qualitative \cite[e.g.][]{gadiraju2023wouldn, mack2024they, qadri2023ai}, the vast majority of research around harms caused is computational \cite{blodgett2020language} with little to no involvement of direct stakeholders: real people who use Stable Diffusion in their daily commercial or personal contexts. Specific to Stable Diffusion, one such exploration is \citet{ghosh2023person}, who computationally demonstrate how Stable Diffusion outputs for a `Person' skew towards light-skinned Western men, almost erase nonbinary and Indigenous identities, and heavily sexualize women of color. While this extensively shows potential harms across a wide range of identities, it ignores the voices of the people whose lives these harms might impact. In this study, we examine perceptions of harms caused by Stable Diffusion outputs as told by such users. 

\subsection{Representational Harms caused by T2Is}\label{subsec:rep-harms}

It is well-established that T2Is such as Stable Diffusion exhibit and embed a set of biases. Having said that, bias is not inherently always bad, and to say that T2Is exhibit bias is not a bad thing. Humans have biases, and therefore it is but natural that human-designed systems embed biases, and in turn apply them in their operations. Such biases are not inherently negative \cite{blodgett2020language, miceli2022studying}, and it is only when they embed unjust stereotypes about people or groups that they can cause harm.

\citet{barocas2017problem} break down such harms into two broad types: allocational and representational. Allocational harms refer to those caused when certain populations are denied access to important opportunities or resources,  whereas representational harms are caused more when depictions of certain groups are unfairly constructed and lead to the formation of incorrect and unfair stereotypes about them which has long-term societal impacts at scale. Representational harms are typically borne disproportionately by traditionally marginalized populations \cite[e.g.,][]{luccioni2023stable, gautam2024melting, gadiraju2023wouldn}. 

Representational harms caused by systems such as T2Is are categorized into five types \cite{dev2021measures}: \textit{stereotyping} or the overrepresentation of some opinions about an identity, \textit{disparagement} or the idea that some groups of people are lesser than others, \textit{dehumanization} or the practice of treating certain groups of people as sub-human, \textit{erasure} or the lack of representation of groups of people, and \textit{quality of service} or models providing unequal outcomes for different groups of people. This taxonomy is also not exhaustive, especially given how novel development and use cases of T2I may result in novel harms. Some examples of T2Is causing harms include \citet{qadri2023ai}'s and \citet{ghosh2024generative}'s works on Orientalized depictions of Indian cultures by T2Is, \citet{mack2024they} showing how T2Is equate `disability' with wheelchairs, and \citet{bianchi2023easily}'s evidencing that T2Is reinforce notions of Whiteness being the default while propagating stereotypes associating people of color with poverty and crime, to name a few. In this study, we further explicate representational harms caused by T2Is. 

\section{Methods}\label{sec:methods}

In this section, we overview prompt formation and generating images with Stable Diffusion, descriptions of the human subjects study, study procedure, and analysis techniques.

\subsection{Prompt Formation and Output Generation}\label{subsec:prompt-image}

From \citet{ghosh2023person}'s study around the defaults of personhood within Stable Diffusion, we adopt the 136 prompts they used. We use their baseline `a front-facing photo of a person' without gender information, and their three gender prompts featuring `man', `woman', and 'person of nonbinary gender'. We also constructed their prompts for 6 continents (Africa, Asia, Oceania, Europe, North America, and Latin America) e.g.,`a front-facing photo of a person from Africa', 27 countries e.g., `a front-facing photo of a person from Indonesia', and 108 combinations of countries/continents and genders e.g., `a front-facing photo of a woman from Bangladesh'. The full set of countries is (in alphabetical order): Argentina, Australia, Bangladesh, Brazil, Canada, China, Colombia, Egypt, Ghana, Ethiopia, France, Germany, India, Indonesia, Italy, Japan, Mexico, New Zealand, Nigeria, Pakistan, Papua New Guinea, Peru, Russia, South Africa, the UK, the USA, and Venezuela. 

We used these to generate 50 images per prompt pursuant to previous research \cite[e.g.,][]{fraser2023friendly,ghosh2023person,mandal2023multimodal}, with the latest free version of Stable Diffusion (v2.1). Similar to \citet{ghosh2023person}, we address prompts in this paper in condensed forms e.g., the prompt `a front-facing photo of a person' is hereafter written as `Person', the prompt `a front-facing photo of a man from India' is written as `Man from India', etc.

\subsection{Participant Recruitment}

To gather user perspectives for Stable Diffusion outputs, we crowdsourced participation using the Prolific platform, based on literature showing its quality over other platforms \cite{prolific}. We based compensation on our local minimum wage of US\$20/hr, with payment disbursed as a function of time taken. We sourced for all genders and nationalities mentioned above, and filtered for English speakers such that they could read the instructions in our web application and prompts. We collected 133 valid responses (see Appendix \ref{app:data}) in October-November 2023.

We also invited Stable Diffusion users to interview with us, recruiting eligible users over 18 years old and fluent in English via a survey inviting them to participate in a 45-60 minute Zoom interview for US\$20 compensation. We conducted 14 interviews (participant information in Table \ref{tab:participants}). 

\subsection{Study Procedures}

\subsubsection{Crowdsourced Study through Visual Elicitation} We used a visual elicitation method known as `Diamond Ranking', a qualitative method utilized to obtain structured data for image comparisons by humans \cite{diamond}. This involves an instrument of cells arranged in a diamond, and asking individuals to place one image per cell from a set of images. Cells are ordered from most (topmost) to least (bottom-most) of some researcher-defined metric. This method was chosen over other visual ranking methods to leverage the bands of the diamond to understand user conceptualizations of groupings of images, as per \citet{lutz_diamond}'s study utilizing Diamond Ranking with TikTok content experiences of marginalized users.

\begin{figure}[h]
  \centering
  \fbox{\includegraphics[width=0.5\linewidth]{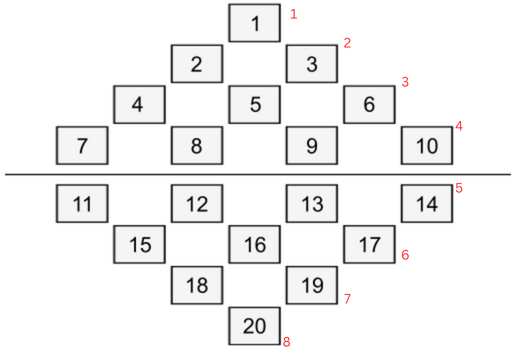}}
  \caption{A sample 20-cell diamond, with cell numbers indicated in black font and row numbers in red font.}
    \label{fig:diamond}
\end{figure}

We developed a web-based Diamond Ranking tool, with the topmost cell indicating highest similarity to participant expectations of a given prompt output and the bottom-most cell indicating least similarity, with images in the same row considered equally similar. We piloted various Diamond arrangements with fellow researchers, concluding that a 20-cell Diamond (shown in Figure \ref{fig:diamond}) would strike the best balance between obtaining a valuable amount of data and not overwhelming participants with too many choices. We also added a horizontal line bisecting the Diamond, to avoid the 8 cells across rows 4 and 5 being valued the same. 

Prolific participants were provided instructions about the Diamond Ranking task in an intro screen and then asked to fill a Diamond for the prompt of `Person', i.e., select 20 images out of a displayed 50 to populate cells, ordering them from most to least similar to own expectations of the Stable Diffusion output for `Person'. 

\subsubsection{Interviews}

\begin{table*}[h]
  \begin{tabular}{p{0.2cm}p{5.3cm}p{4.3cm}p{6.5cm}}
    \toprule
    ID & Self-Reported Gender and Nationality & Stable Diffusion Prompt 1 & Stable Diffusion Prompt 2 \\
    \midrule
    P1 & Man, and Canadian & `Person from Canada' & `Man from Canada'\\
    P2 & Woman, and Chinese & `Person from China' & `Woman from China'\\
    P3 & Woman, and American & `Person from USA' & `Woman from USA'\\
    P4 & Nonbinary Gender, and Indian & `Person of Nonbinary Gender' & `Person of Nonbinary Gender from India'\\
    P5 & Woman, and British & `Person from UK' & `Woman from UK'\\
    P6 & Woman, and Mexican & `Person from Mexico' & `Woman from Mexico'\\
    P7 & Man, and Indian & `Person from India' & `Man from India'\\
    P8 & Man, and Argentinian & `Person from Argentina' & `Man from Argentina'\\
    P9 & Woman, and Egyptian & `Person from Egypt' & `Woman from Egypt'\\
    P10 & Woman, and Venezuelan & `Person from Venezuela' & `Woman from Venezuela'\\
    P11 & Woman, and Pakistani & `Person from Pakistan' & `Woman from Pakistan'\\
    P12 & Woman, and Bangladeshi & `Person from Bangladesh' & `Woman from Bangladesh'\\
    P13 & Nonbinary Gender, and British & `Person of Nonbinary Gender' & `Person of Nonbinary Gender from UK'\\
    P14 & Nonbinary Gender, and Australian & `Person of Nonbinary Gender' & `Person of Nonbinary Gender from Australia'\\
  \bottomrule
\end{tabular}
\caption{Interviewee Information, showing Participant ID, self-reported gender and nationality, and the two prompts for which Stable Diffusion outputs were shown to them.}
\label{tab:participants}
\end{table*}

We supplement our findings from crowdsourced data with interviews of Stable Diffusion users around their experiences of their identities being representations by Stable Diffusion. Interviews began with soliciting consent to record, proceeding with questions about participants' experiences with Stable Diffusion and, if they had experience generating human faces, their thoughts on such outputs. The second component of the interview involved the use of the Diamond Ranking Tool. Based on participant survey responses of self-reported gender/nationality, we offered them 2 prompts out of a permutation of 3, e.g., if a participant identified as male and Indian, we asked them to select 2 prompts out of `Man from India', `Person from India', and `Man'. Having selected two prompts, we then conducted two sessions with the Diamond Ranking Tool, asking them to fill the diamonds for their prompts. We also asked about the rationale behind their choices, and whether they saw themselves represented within the images. The study was approved by our university's IRB.

\subsection{Analysis Techniques}\label{subsec:analysis}

We analyzed crowdsourced data across 133 Diamonds with images from the Stable Diffusion output for `Person'. We counted frequencies of each image being placed in each level of Diamonds (see Figure \ref{fig:diamond}), and built an overall ranking for images considered most to least similar to participant expectations of Stable Diffusion output for `Person'.

We performed a computational analysis of the 50 Stable Diffusion images generated for `Person', as measured through the \textit{cosine similarity} metric: an approach to measure how similar two input vectors are, with scores ranging from 0 (very dissimilar) to 1 (identical) \cite{singhal2001modern}. We perform cosine similarity comparisons for images pursuant to a well-established approach within the field \cite[e.g.][]{ghosh2023person, sejal2016image, wolfe2022markedness}. We measure \textit{intra-set cosine similarities} within sets of 50 outputs per prompts by comparing how each image is similar to the other 49 in the same set, and report an average similarity score, such that we can identify which image Stable Diffusion considers closest to the given prompt. E.g., for the 50 Stable Diffusion outputs of `Person', we compare every image to each of the other 49 via cosine similarity, compute a mean cosine similarity score, and arrange them in ranked order. To determine trends between user expectations of Stable Diffusion outputs for `Person' and generated results, we examined how most and least popularly-selected images compared with intra-set cosine similarity ranks. 

We performed thematic analysis \cite{braun2006using} to analyze our interviews, and provide salient codes in Table \ref{tab:codes}. We also performed further cosine similarity comparison tasks to augment some of our interview findings. In Section \ref{subsec:erasure}, we perform intra-set cosine similarities individually across two sets of 50 images for two given prompts. For Section \ref{subsec:dehuman}, we verify interview observations by comparing results from a given prompt with Stable Diffusion outputs for `Woman'. In this case, we examined the average similarity of each image from results for the prompt used in Section \ref{subsec:dehuman} with each of the 50 Stable Diffusion outputs for `Woman', to produce an overall cosine similarity score.

We feature in this study no more than 1-2 participants per country and do not make any claims of their comments being representative of their cultures, honoring the feminist research tradition of situating comments in lived experiences and expertises \cite{haraway1988situated}. Such generalized claims would perform the same homogenization that we (in Section \ref{subsec:erasure}) identify within Stable Diffusion, and place an unfair burden upon interviewees to be perfect spokespeople for their cultures, a burden often assigned disproportionately to traditionally marginalized peoples. 

\section{Findings}
We demonstrate gaps between user perspectives of `Person' and Stable Diffusion outputs, and highlight how users experience representational harms \cite{barocas2017problem}, with the most prominent ones being erasure, stereotyping, dehumanization, and disparagement \cite{dev2021measures}. Quotes are anonymized to avoid revealing identifiable information. 

\subsection{Comparing User Perspectives of `Person' and Stable Diffusion Outputs}\label{subsec:findings-person}

Among 133 Prolific participants, each Stable Diffusion output for `Person' appeared at least 26 times and at most 83 times across the diamonds. The intra-set cosine similarity scores for this set of images ranged within 0.83-0.67, indicating variation in results for the same prompt.

Cross-referencing the data from user selections with intra-set cosine-similarity scores, we observe a gap between the two. Of the top 5 highest intra-set cosine-similarity scores (ranging 0.83-0.80), 4 were selected within the Diamonds only 31-39 times and most commonly within level 5 of the diamonds (see Figure \ref{fig:diamond} for information about levels). This indicates that the most central depiction of a `Person' by Stable Diffusion are only slightly in line with user expectations. However, most important to note within these is an image which ranks 3\textsuperscript{rd} within intra-set cosine-similarity scores (with a score of 0.81) but is found 74 times within Prolific user diamonds and most commonly in row 7 (16 times) and row 8 (39 times), implying that it was consistently highly dissimilar to user expectations and indicative of a large disconnect between such and generated outputs. 

The images which appear the most consistently in rows 1 and 2 across user-completed Diamonds rank near the bottom of intra-set cosine-similarity scores. The image chosen the most within rows 1 and 2, appearing 14 times in row 1 and 5 times in row 2, has an intra-set cosine-similarity score of 0.73 and ranks 40\textsuperscript{th}. The next highest frequency across rows 1 and 2 is of an image that users place in these rows 18 times -- 5 times in row 1 and 13 times in row 2 -- but also has an intra-set cosine-similarity score of 0.73, ranking 38\textsuperscript{th}. It is not until we reach the image that occurs the 7\textsuperscript{th} most in rows 1 and 2 that we observe an intra-set cosine-similarity ranking in the top 25, with the score of 0.78 ranking 14\textsuperscript{th}. This gulf in the most commonly user-selected images being rated as not too similar to other Stable Diffusion outputs for `Person' is further evidence of how far apart user expectations of Stable Diffusion outputs are from those actually generated.

Building on this disconnect, we further show patterns of Stable Diffusion users feeling underrepresented, dissatisfied and angered at the depiction of their own identities within outputs, which cause the representational harms of erasure, stereotyping, disparagement, and dehumanization. 

\subsection{Erasure and Stereotyping within Stable Diffusion Outputs}\label{subsec:erasure}

\begin{figure*}[t]
\centering
\begin{subfigure}[t]{0.22\textwidth}
    \fbox{\includegraphics[width=\textwidth]{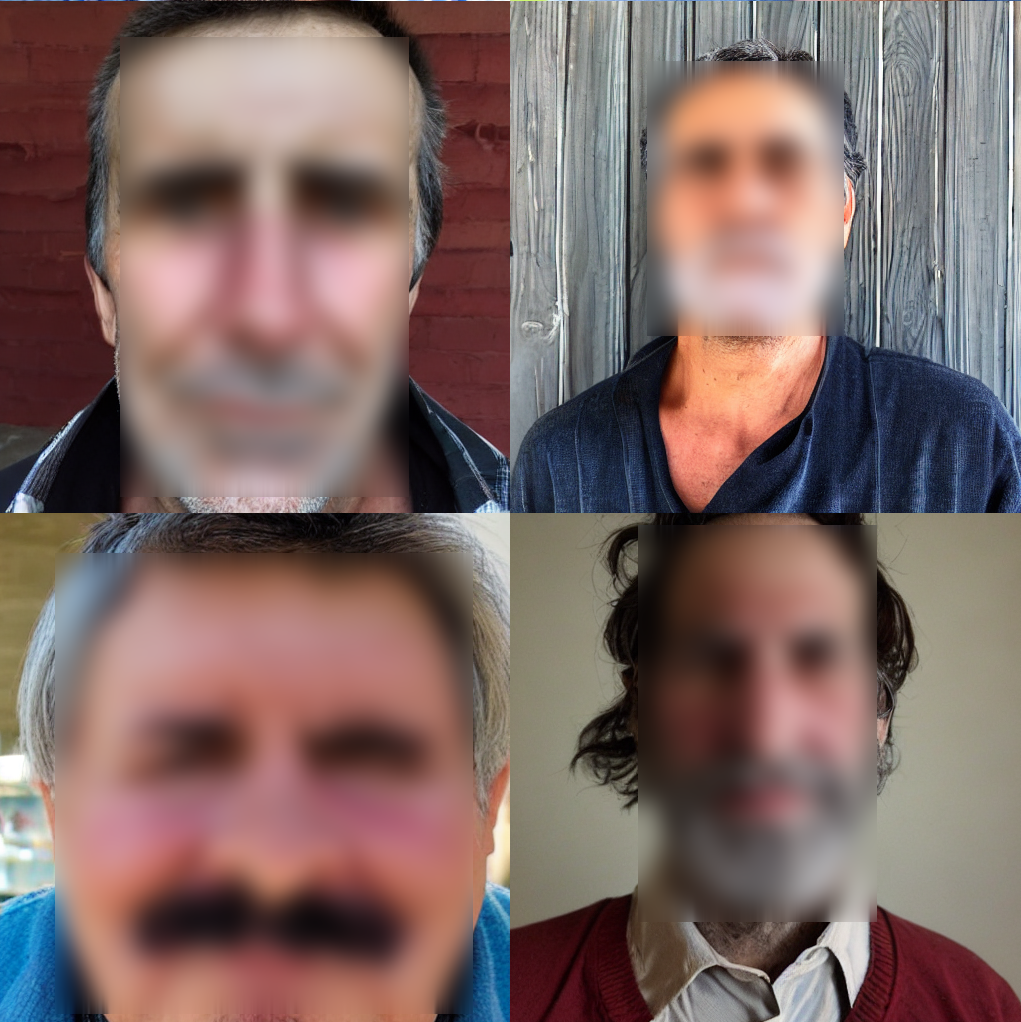}}
    \caption{Output for `Person from \\ Argentina', as shown to P8.}
    \label{fig:argentina}
\end{subfigure}
\hspace{0.5em}
\begin{subfigure}[t]{0.22\textwidth}
    \fbox{\includegraphics[width=\textwidth]{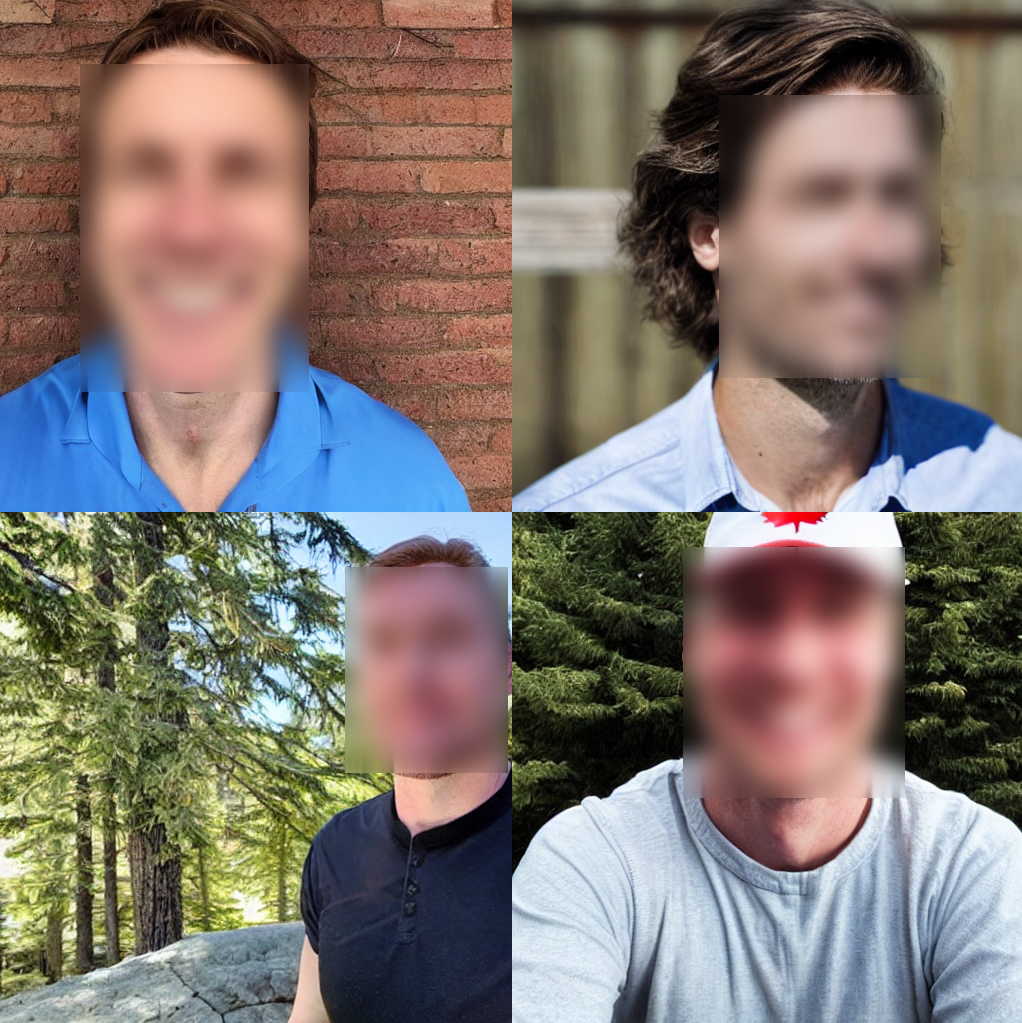}}
    \caption{Output for `Man from Canada', as shown to P1.}
    \label{fig:canada}
\end{subfigure}
\hspace{0.5em}
\begin{subfigure}[t]{0.22\textwidth}
    \fbox{\includegraphics[width=\textwidth]{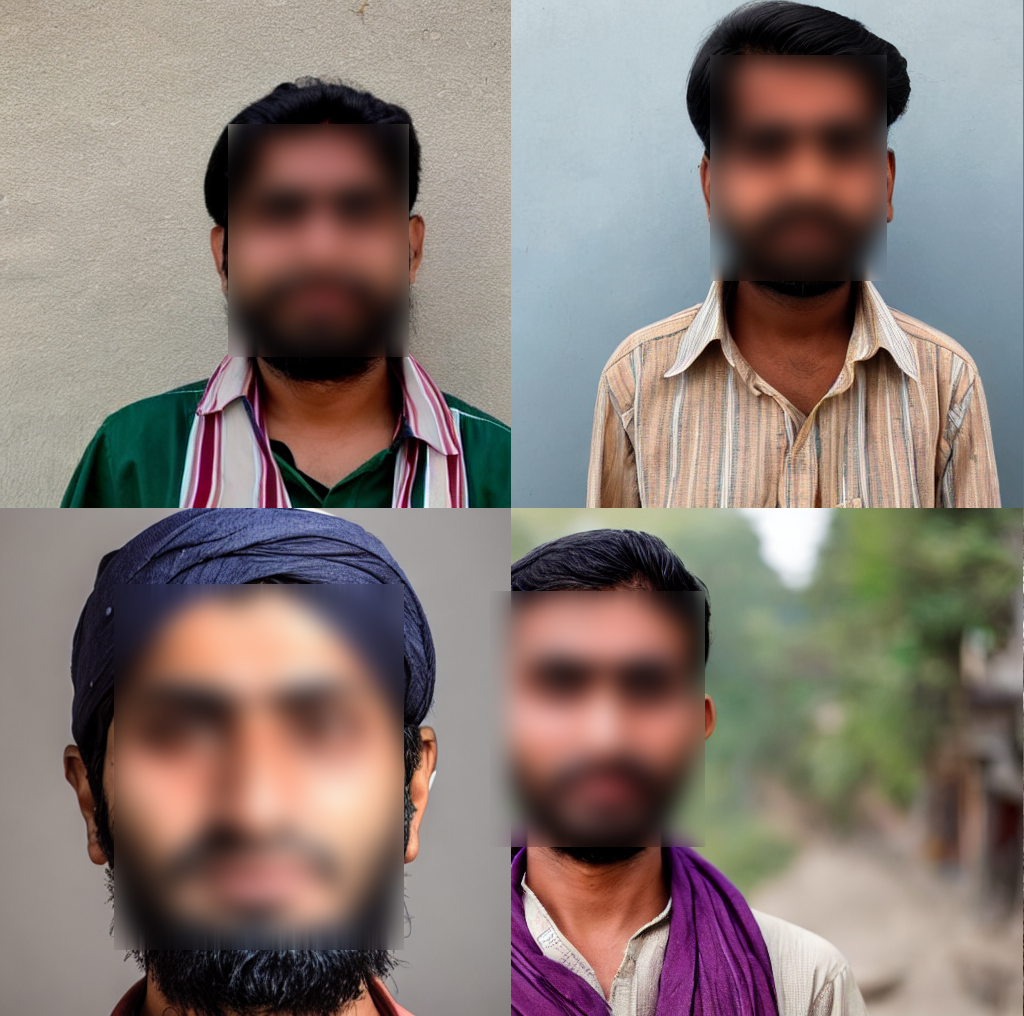}}
    \caption{Output for `Person from Pakistan', as shown to P11.}
    \label{fig:pakistan}
\end{subfigure}
\hspace{0.5em}
\begin{subfigure}[t]{0.22\textwidth}
    \fbox{\includegraphics[width=\textwidth]{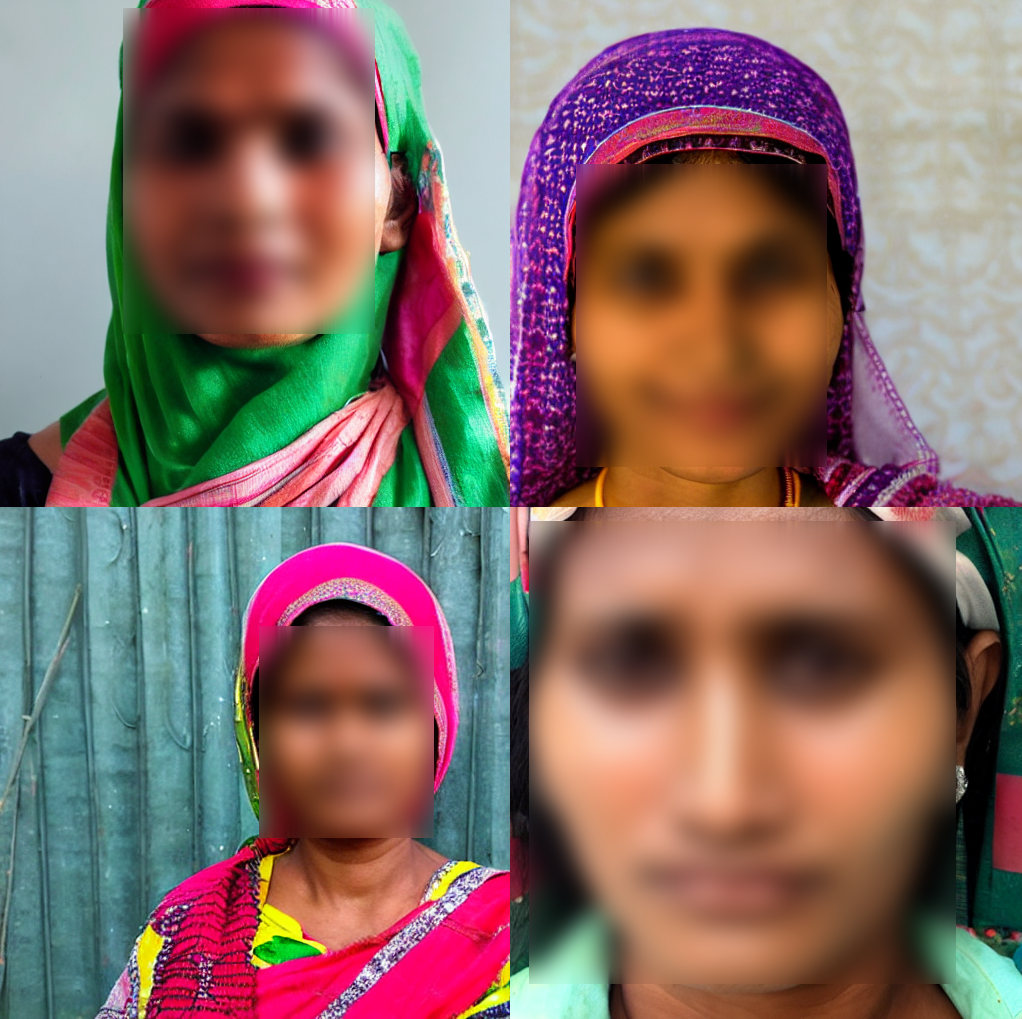}}
    \caption{Output for `Woman from Bangladesh', as shown to P12.}
    \label{fig:bangladesh}
\end{subfigure}
\\
\begin{subfigure}[t]{0.22\textwidth}
    \fbox{\includegraphics[width=\textwidth]{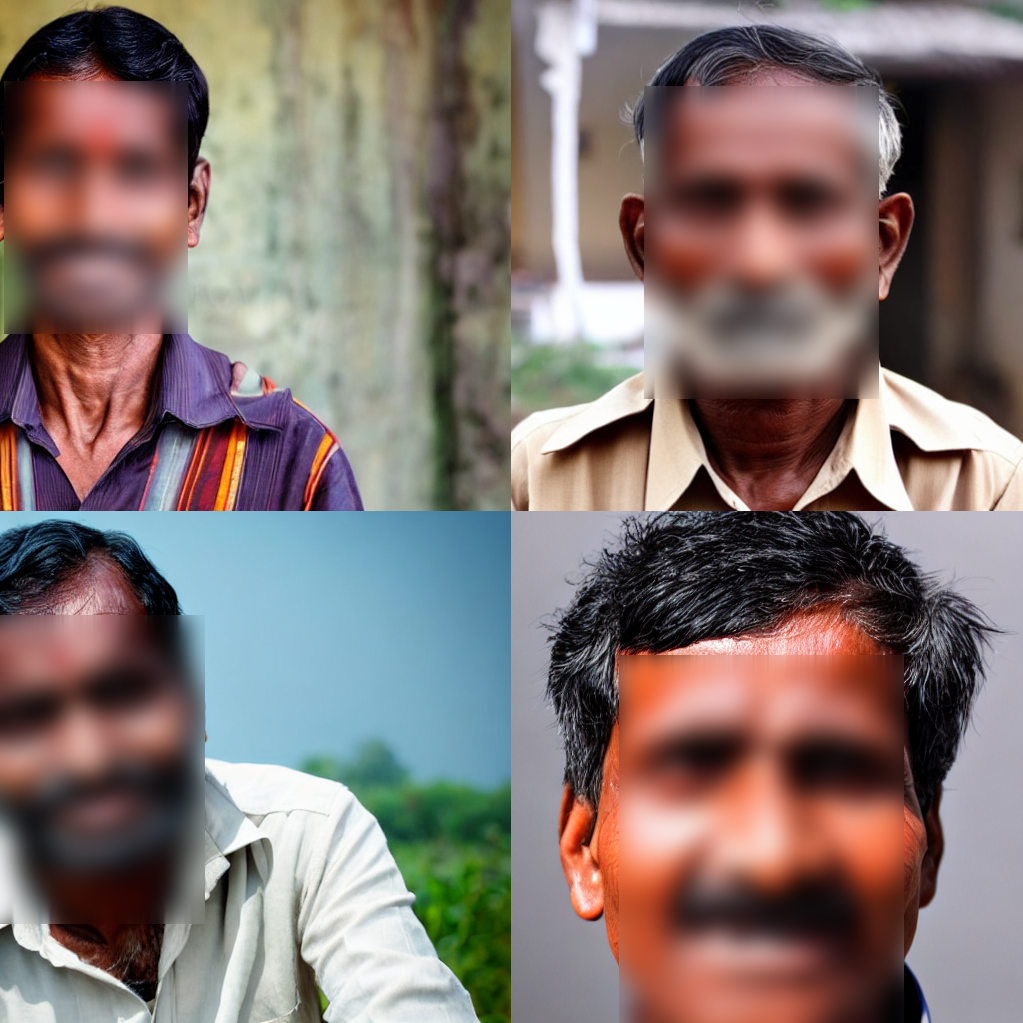}}
    \caption{Output for `Person from \\India', as shown to P7.}
    \label{fig:india}
\end{subfigure}
\hspace{0.5em}
\begin{subfigure}[t]{0.22\textwidth}
    \fbox{\includegraphics[width=\textwidth]{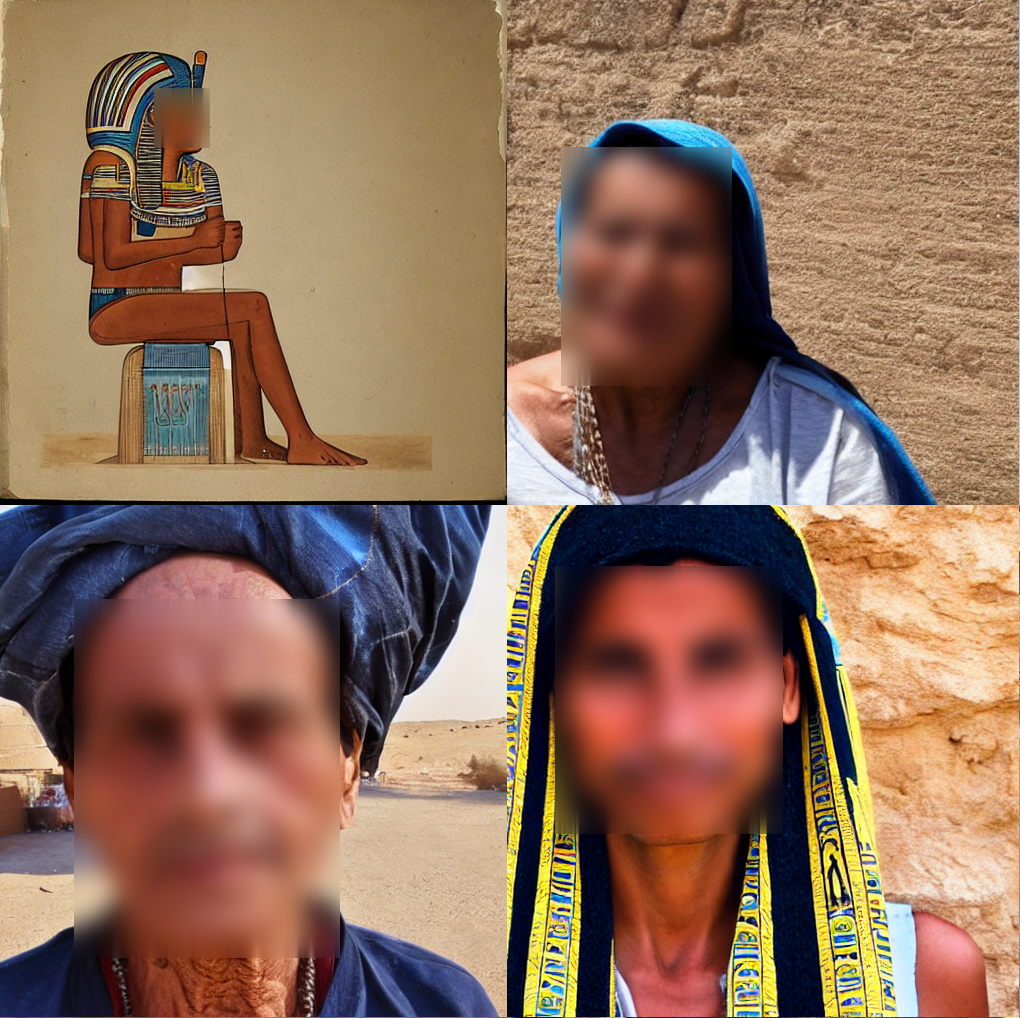}}
    \caption{Output for `Person from Egypt', as shown to P9.}
    \label{fig:egypt}
\end{subfigure}
\hspace{0.5em}
\begin{subfigure}[t]{0.22\textwidth}
    \fbox{\includegraphics[width=\textwidth]{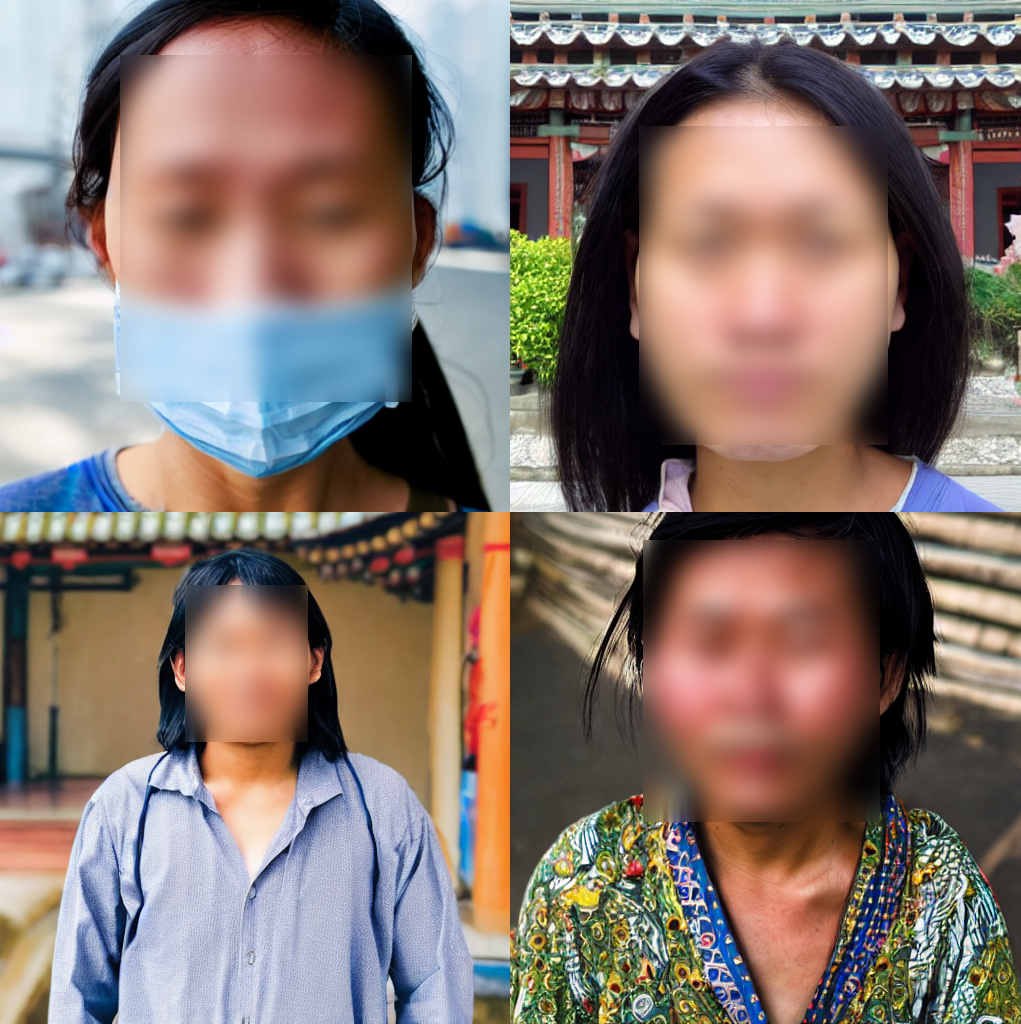}}
    \caption{Output for `Person from China', as shown to P2.}
    \label{fig:china}
\end{subfigure}
\hspace{0.5em}
\begin{subfigure}[t]{0.22\textwidth}
    \fbox{\includegraphics[width=\textwidth]{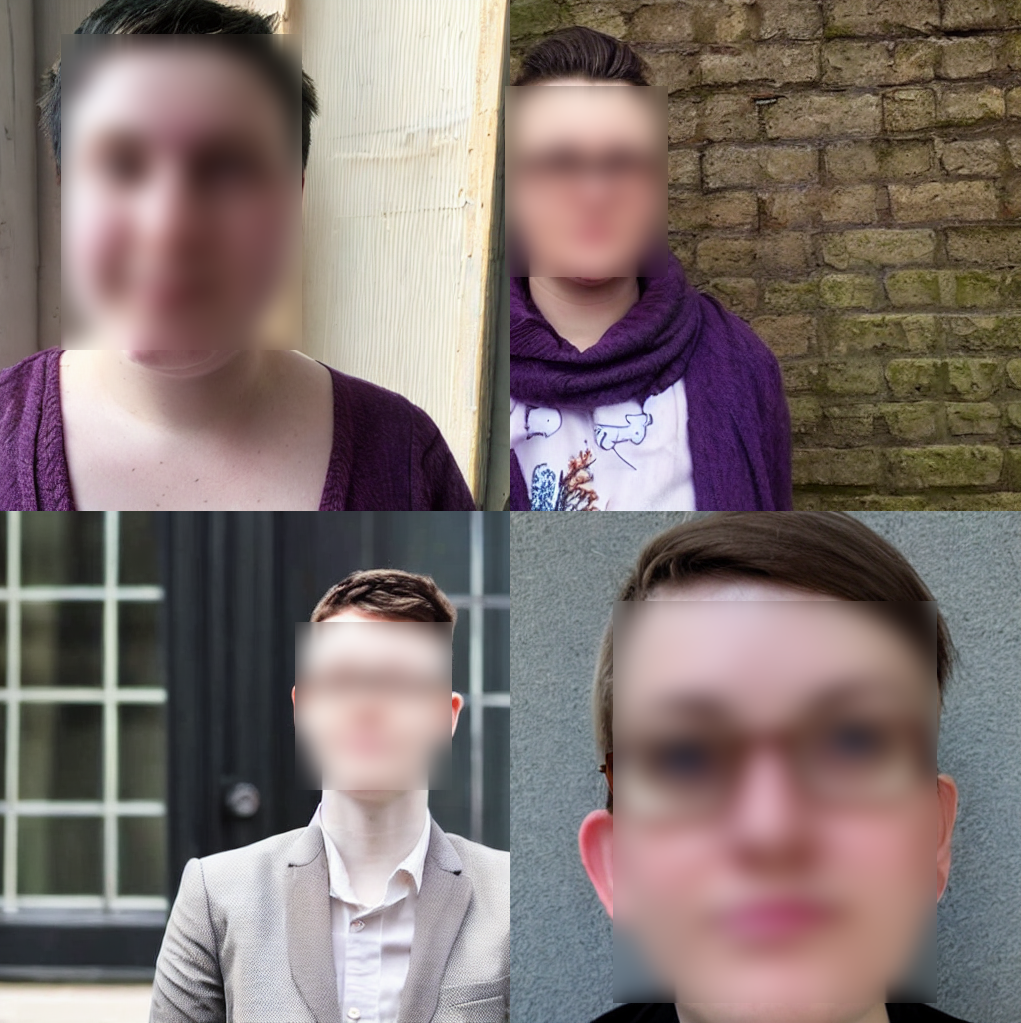}}
    \caption{Output for `Person of Nonbinary Gender from the UK', shown to P13.}
    \label{fig:enby-uk}
\end{subfigure}
\caption{Illustrative examples of Stable Diffusion outputs showing Stereotyping and Erasure, in 2x2 grids.}
\label{fig:dehumanization}
\end{figure*}
Within T2I outputs, stereotyping is the representational harm of outputs containing attributes based on overgeneralized and often negative beliefs about identities, while erasure occurs when the propagation of such stereotypes leaves little to no room for representing other aspects of such identities \cite{dev2021measures, katzman2023taxonomizing}. These harms prominently exist within our findings, as participants found themselves underrepresented within Stable Diffusion outputs of gender and nationalities they identified with.

A significant example of \textbf{erasure} was around gender, as participants identifying as women saw minimal representation of themselves in Stable Diffusion outputs of `Person from X', where X is a country in their prompts. P5, a woman from the UK, mentioned how \textit{``when [they] heard [the interviewer] say `person', [they] expected to see somewhat of a split between men and women, but [the output] is mostly men."} A woman from Bangladesh, P12 was struck by disbelief as they asked \textit{``the one you're showing me here is person? Not man? I'm literally not seeing any women here.} P5 and P12 were surprised to see how underrepresented women were in the outputs for `Person from the UK' and `Person from Bangladesh', thus providing empirical evidence to supplement a previously computationally demonstrated finding on the masculine default \cite{ghosh2023person}.

For some, the lack of representation of their identities emerged as Stable Diffusion depicted homogenized characteristics of prompted countries. An Argentinian man, P8 felt the outputs for `Person from Argentina' and `Man from Argentina' (Figure \ref{fig:argentina}) did their country a disservice since \textit{``Argentina is so much more diverse than what is shown here. We have black and brown people, as well as people who look nothing like any of these, but I'm not seeing them at all here."} This homogenization into a national default was further jarring for P1, a man from Canada of self-identified Asian descent, who saw outputs for both `Person from Canada' and `Man from Canada' (Figure \ref{fig:canada}) not contain a single image they personally identified with. They remarked \textit{``Canada is no longer a country of tall white guys with beards, and has not been for a long time. There is a lot of diversity here, a lot of people who look like me, and to see that diversity completely erased here is pretty sad."} P1 elaborated that while these attributes might have been ubiquitous in Canada in the past, that has since changed, and the erasure of this change within depictions of Canadian identities was disappointing. 

The harm of \textbf{stereotyping} was also observed by participants from the South Asian countries of Bangladesh (P10), India (P4 and P7), and Pakistan (P11). All these participants noted the stereotype of `South Asia as impoverished and under-developed' \cite{qadri2023ai} in their respective findings. P11 noted how outputs for `Person from Pakistan' (Figure \ref{fig:pakistan}) were such that \textit{`` these [images] look like people from rural Pakistan, in fact almost all of them do,"} and P12 observed the outputs from `Person from Bangladesh' and `Woman from Bangladesh' (Figure \ref{fig:bangladesh}) as \textit{``the stereotypical representations of village people in Bangladesh."}

It is important to note a novel nuance within this pattern of depicting South Asian identities as `impoverished and under-developed' \cite{qadri2023ai}, which was best explained by P7. A man from India, P7 too observed the prominence of \textit{``rural lifestyles"} within the results for `Person from India' (Figure \ref{fig:india}) and `Man from India', which they did not identify with, but also pointed out that \textit{``this result makes sense, because [they are] not what is considered the `majority' population in India."} As a self-identified person from an urban city, P7 understood that in a country where the majority population is rural \cite{india_stats}, their identity not being represented by a service which theoretically operates on majority patterns made sense. However, the harm of erasure still applies, because across a set of 50 results each for 2 prompts, P7 \textit{``still expected to see a little bit of representation, maybe in 1-2 images."} Stable Diffusion thus did a poor job in representing their identity even when they gave it the grace of showing a few or even one image that represents them, and they were only disappointed when they were afforded absolutely no representation. 

Perhaps the most heartbreaking experience of erasure came from P12, a woman from Bangladesh. After their disappointment at seeing an absence of women in the outputs for `Person from Bangladesh', they expected better representation of themselves in the outputs for `Woman from Bangladesh'. While the latter (Figure \ref{fig:bangladesh}) did offer some representation of their womanhood and they found the images to be pretty accurate in their own experience, P12 still felt erased within these images. They identified as part of an Indigenous community within Bangladesh that has been fighting for and gaining representation, and to see absolutely no hints of representation within these outputs of themselves and their community was upsetting. In their own words, \textit{``We're small but we're there...I understand that I'm a minority in my country, but the fact that Stable Diffusion shows no images that look like me in a set of 50 is disappointing.}

The propagation of stereotypes about countries continued across our findings, as we saw the prevalence of both ancient and modern stereotypes being depicted. P9, a woman from Egypt, noted how the outputs for `Person from Egypt' (Figure \ref{fig:egypt}) and `Woman from Egypt' contained stereotypical features prominent in the ancient days of the Egyptian kingdoms of many thousands of years ago. They remarked that \textit{``Egypt is a lot more than sands, pyramids and sphinxes. Stable Diffusion seems like its stuck thousands of years ago and thinks of Egypt only in Pharaoh terms."} On the other end of the temporal spectrum, P2, a woman from China, noted how representations of `Person from China' (Figure \ref{fig:china}) and `Woman from China' contained images where \textit{``The people here are wearing masks, which is definitely an effect of Covid."} This leaned in to the stereotype of associating the COVID-19 virus and mask-wearing with China, a stereotype which underpins many hate crimes directed towards Chinese and Asian people over the past few years \cite{gray2021did}, making these results particularly alarming. 

The stereotypical homogenization of identities and resultant erasure of diversity was also prominent where Stable Diffusion outputs were only showing small variations of one or a few baseline constructions of faces or features. This was first identified by P10, a woman from Venezuela, who mentioned how a vast majority of images were \textit{``Very very similar to each other. They all have the same nose!"}. A similar sentiment was also raised by P13 around the output for `Person of Nonbinary Gender from the UK' as they observed that \textit{``A lot of these people look literally the exact same as each other. There's genuinely no difference."} This similarity across a set of 50 images was unexpected to P10 and P13, since it would be analogously unexpected for 50 real people. 

\begin{figure*}[t]
\hspace{0.5em}
\begin{subfigure}[t]{0.22\textwidth}
    \fbox{\includegraphics[width=\textwidth]{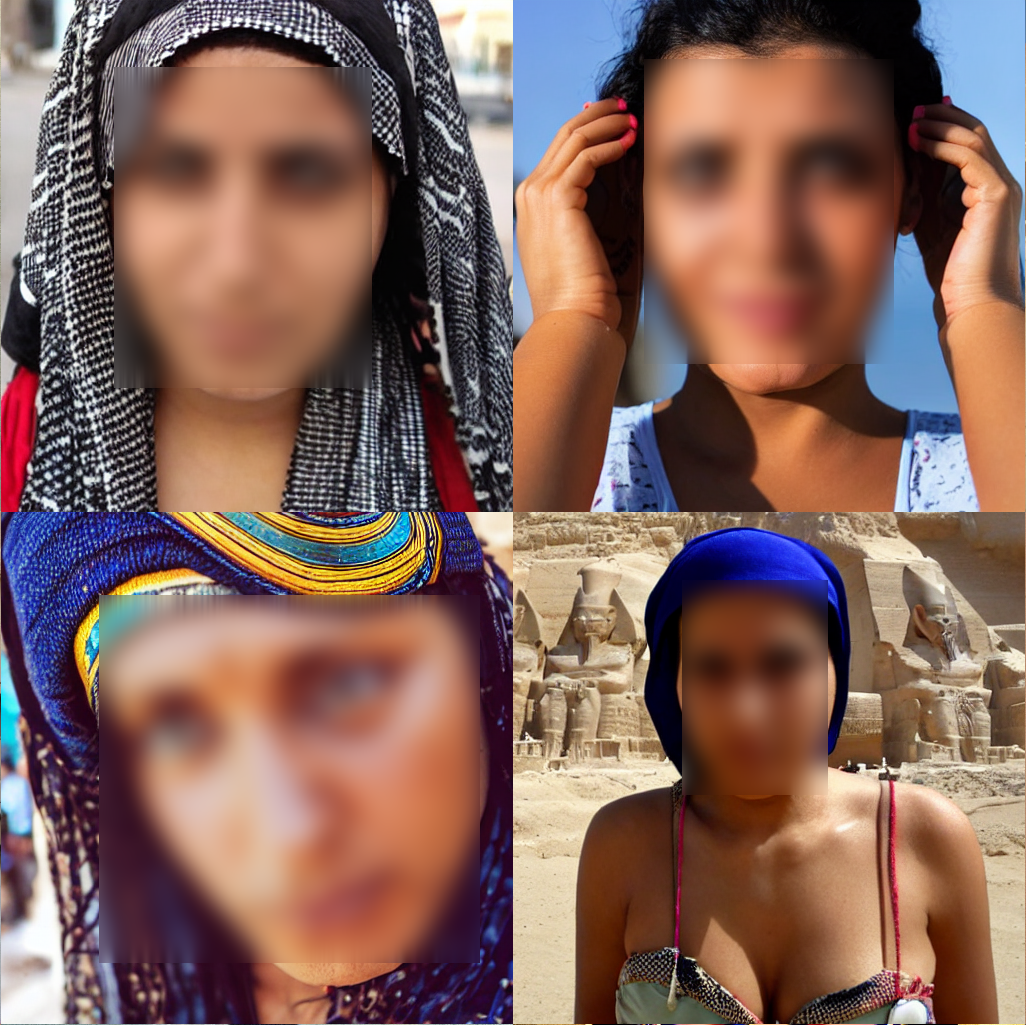}}
    \caption{Output for `Woman from Egypt', shown to P9.}
    \label{fig:egypt-woman}
\end{subfigure}
\hspace{0.5em}
\begin{subfigure}[t]{0.22\textwidth}
    \fbox{\includegraphics[width=\textwidth]{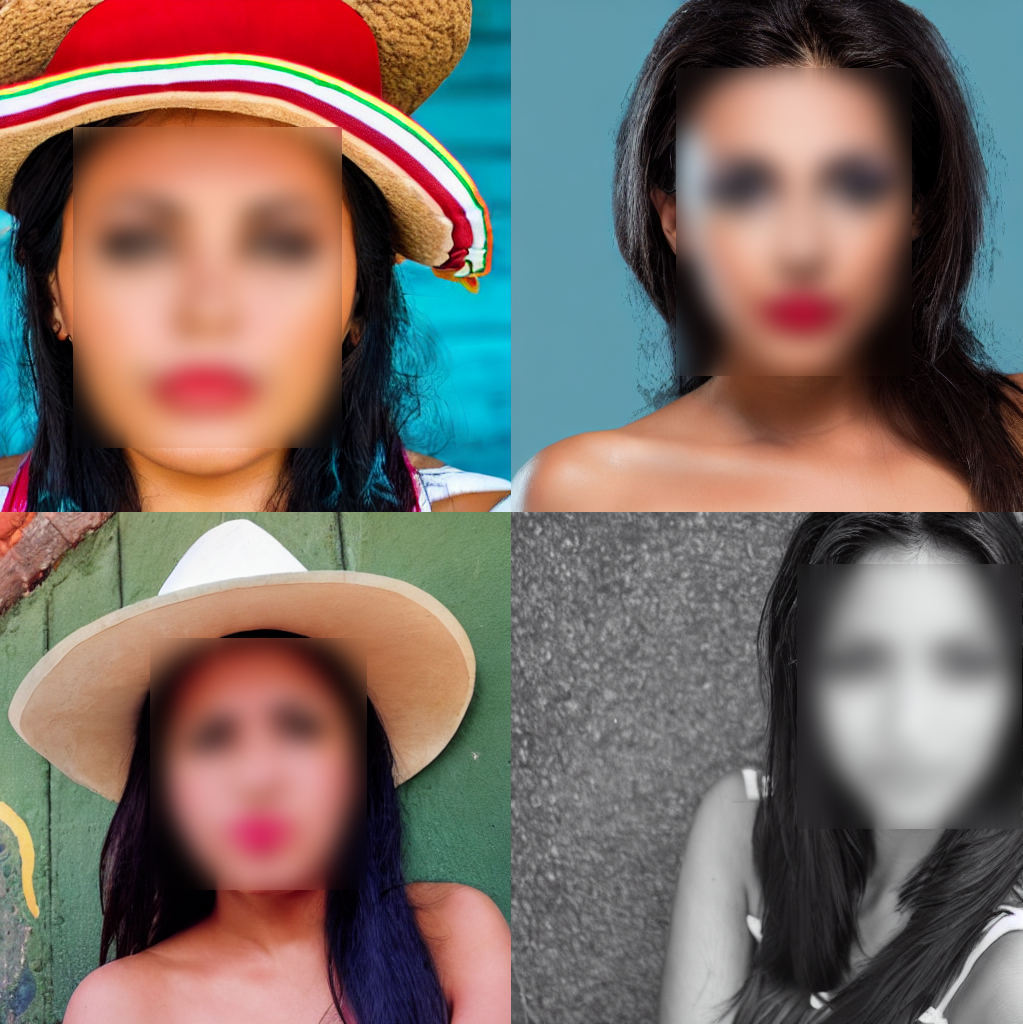}}
    \caption{Output for `Woman from Mexico', shown to P6.}
    \label{fig:mexico}
\end{subfigure}
\hspace{0.5em}
\begin{subfigure}[t]{0.22\textwidth}
    \fbox{\includegraphics[width=\textwidth]{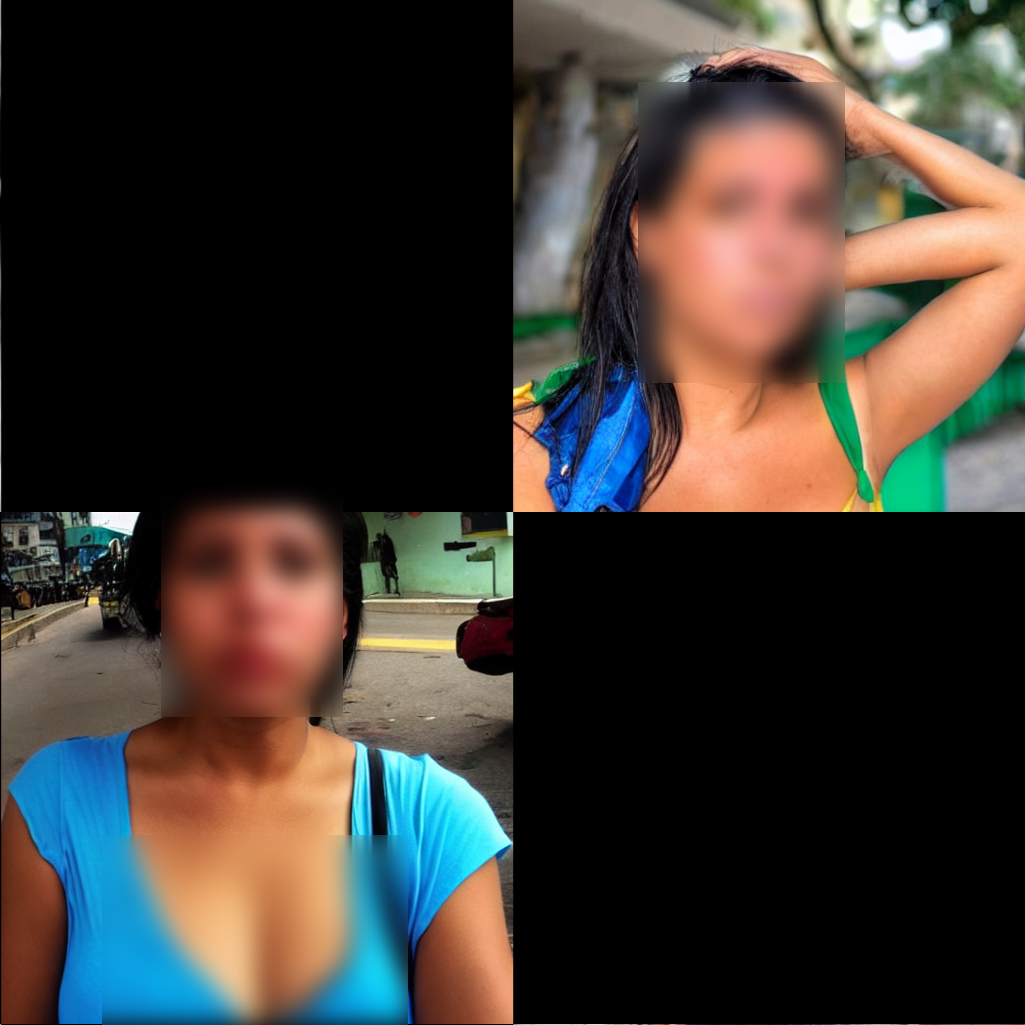}}
    \caption{Output for `Woman from Venezuela', shown to P10.}
    \label{fig:venezuela}
\end{subfigure}
\hspace{0.5em}
\begin{subfigure}[t]{0.22\textwidth}
    \fbox{\includegraphics[width=\textwidth]{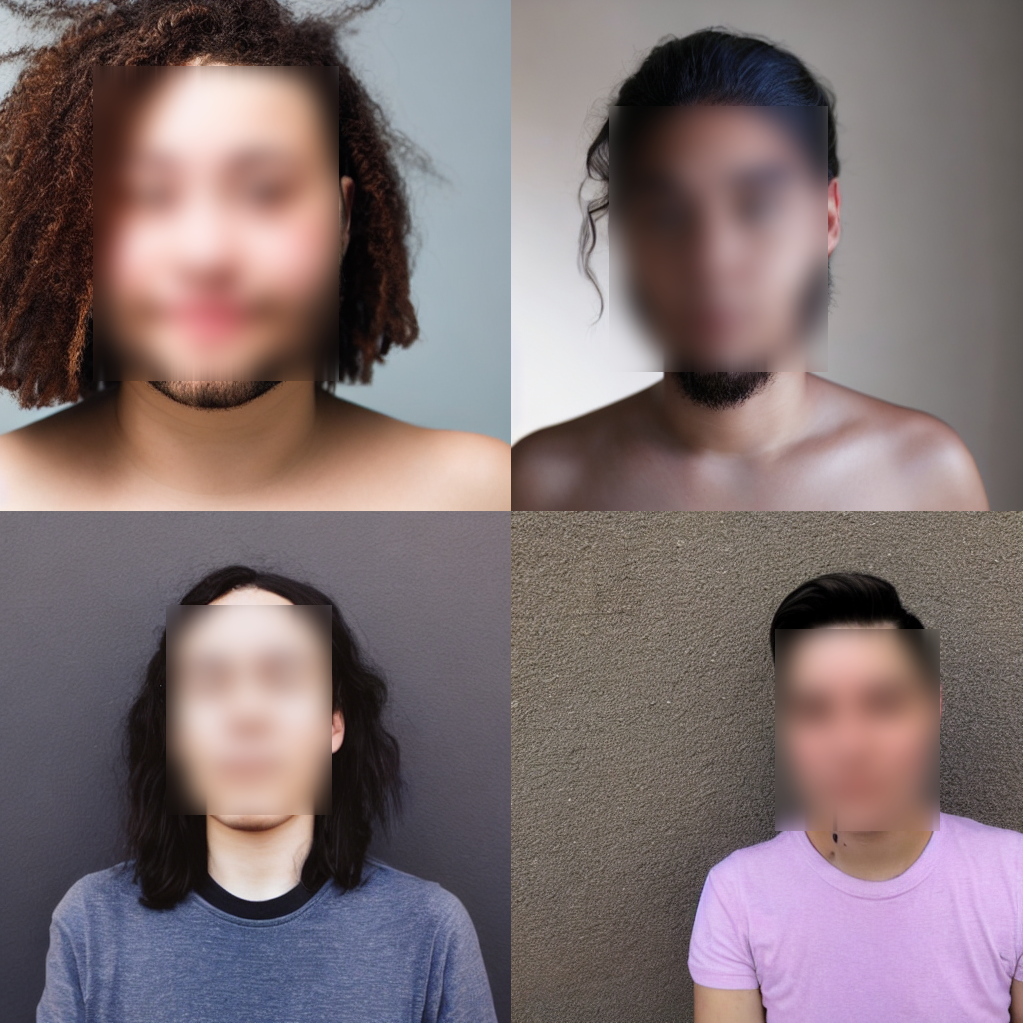}}
    \caption{Output for `Person of Nonbinary Gender', shown to P4, P13 and P14.}
    \label{fig:enby}
\end{subfigure}
\caption{Illustrative examples of Stable Diffusion outputs showing Dehumanization, in 2x2 grids.}
\label{fig:dehumanization}
\end{figure*}

We verified this observation of similarity across images by examining the intra-set cosine similarity for `Woman from Venezuela' (Figure \ref{fig:venezuela}) and `Person of Nonbinary Gender from the UK' (Figure \ref{fig:enby-uk}) as described in Section \ref{subsec:analysis}, the ones for which P10 and P13 made the aforementioned comments. For outputs from `Woman from Venezuela', intra-set cosine similarity scores were in the 0.79-0.62 range, while for `Person of Nonbinary Gender from the UK', they fell in the range 0.77-0.60. These scores confirm P10 and P13's observations, because they indicate that images within each set of 50 outputs are highly similar to the others in their sets. Combined with the fact that intra-set cosine similarity scores for `Person' range between 0.83-0.67 (Section \ref{subsec:findings-person}), we confirm how Stable Diffusion homogenizes identities and fails to produce diverse results. 

We thus observe prominent patterns of stereotyping and erasure within Stable Diffusion outputs, as told by participants who expected to be represented in these outputs.

\subsection{Disparagement and Dehumanization within Stable Diffusion Outputs}\label{subsec:dehuman}

\citet{dev2021measures} define the representational harm of disparagement as behaviors reinforcing the notion of some identities being less worthy of respect than others, and contributing to dehumanization or the depiction of such identities as sub-human and Othered. We document evidence of both.  

One of the most salient examples of \textbf{dehumanization} is reflected within the experiences of three of our participants, whose engagement with shown Stable Diffusion outputs were different than others: P6, P9, and P10, who self-identified as women from Mexico, Egypt, and Venezuela respectively. In each case, the results for the prompts `Person from X Country', where X indicates the aforementioned 3 countries, the respective sets of Stable Diffusion outputs all contained at least one black image out of the 50, which in Stable Diffusion's context means that it deemed the output created to be too not-safe-for-work (NSFW) to display and instead renders a black square \cite{ghosh2023person}. Each participant immediately noticed the black square, and asked about them prior to the start of the Diamond Ranking process. All of them were angered to learn that black squares in this context implied that Stable Diffusion self-censored images as NSFW \cite{ghosh2023person}, more so when told that this was a pattern present only in a handful of outputs. P9, from Egypt, considered the prevalence of sexualized images in outputs for `Woman from Egypt' (Figure \ref{fig:egypt-woman}) \textit{``pretty upsetting,"} and asked \textit{``Why would Stable Diffusion choose only certain prompts to give sexualized images??"} P6, a woman from Mexico, noted that even though they only got 1/50 black squares for `Woman from Mexico' (Figure \ref{fig:mexico}), \textit{``some of these pictures are still pretty sexualized, look at the ones that are topless or wearing small clothes, and it's still pretty bad even after the censoring."} 

It is particularly important to note P10's outrage at their slate of results, especially for `Woman from Venezuela' (Figure \ref{fig:venezuela}). This set of results contained 8/50 black squares, which enraged them enough to spend the first few minutes of their Diamond Ranking task being \textit{``too mad to talk."} When they did proceed with the ranking, they also remarked how \textit{``Some of these women aren't wearing any clothes or like making sexy or flirting eyes, and it's so messed up because obviously women in Venezuela don't just walk around like this."} While the finding of women of color, particularly Latin American women, being dehumanized as overtly sexual by Stable Diffusion isn't novel and has been discussed by \citet{ghosh2023person} as a result of decades of media representations sexualizing Latina bodies \cite[e.g.,][]{molina2010dangerous, fitch2009side, guzman2004brain, mensa2022portraying}, our findings offer the grim reality of how users who identify with these groups perceive the harms hitherto only computationally demonstrated. The anger felt by P10 is one of the sharpest documentations of representational harms within our paper. 

The harm of \textbf{disparagement} occurs when Stable Diffusion tries to render outputs of non-binary individuals, as evidenced by the experiences of P4, P13, and P14, all nonbinary people. These interviewees stated that not only did Stable Diffusion outputs of nonbinary people (e.g., as captured in `Person of Nonbinary Gender') miss the mark in terms of what they were hoping, but also that the outputs (Figure \ref{fig:enby}) seemed to portray them as what P13 called \textit{``women-lite"}. P13 further elaborated on how it \textit{``[felt] like all of these people are women-lite: the faces here look like these were all people assigned female at birth and that is just so wrong, to build that stereotype."} A similar sentiment was shared by P14, who found the images to \textit{``just like base femme faces with masc features superimposed on them."} 

This observation by P13 was interesting, and we decided to supplement this with our own evaluation. We used the previously generated (see Section \ref{subsec:prompt-image}) 50 Stable Diffusion outputs for `Woman' and compared each of those to images for `Person of Nonbinary Gender', for whose outputs P13 made this observation. Based on cosine similarity scores comparing outputs of `Person of Nonbinary Gender' to `Woman', we observed scores falling between 0.78-0.64. Comparing `Person of Nonbinary Gender' to `Man' yielded cosine similarity scores between 0.46-0.38, providing a trend of nonbinary faces being closer to women than men. 

Our interviewees were troubled that Stable Diffusion disparaged them by propagating reductive ideas of nonbinaryness as strongly intertwined with femininity, implying the idea of feminine (or even masculine) identities as the `default' from which they have to change to become nonbinary. Participants such as P4 expressed concerns about \textit{``socially-constructed gender being codified"} by T2Is such as Stable Diffusion, which is also true of other T2Is that portray nonbinary individuals with short lilac/purple hair \cite{wired}, and ended with the hope that someday, T2I outputs would show nonbinary identities as \textit{``normal people."} 

Therefore, our interviews reveal the prospect of Stable Diffusion outputs causing the representational harms of stereotyping, erasure, dehumanization, and disparagement in a variety of ways. Such harms are also shown to be caused more to users with one or more traditionally marginalized identities, e.g., women or people of nonbinary gender, and people from Latin America, Africa or Asia. It is clear that Stable Diffusion as we know it today is deeply flawed in how they represent people, and in severe need of an overhaul. 

\section{Discussion}

Across our findings crowdsourced from Prolific and user interviews, we observe a deep disconnect between user expectations of what Stable Diffusion would produce for a given prompt, and what it actually outputs. In the Prolific findings, we see Stable Diffusion outputs strongly differing from what its global user base expects for a `Person', confirmed across 133 individual users (Section \ref{subsec:findings-person}). This dissatisfaction rises to representational harm \cite{barocas2017problem} across 14 semi-structured interviews as participants found their own genders/nationalities greatly misrepresented by Stable Diffusion, and expressed sentiments ranging from anger to dissatisfaction with such outputs. We detail the occurrence of multiple representational harms through the propagation of stereotypical defaults of nationalities and genders homogenized around a few specific features with the erasure of other diverse forms of expression, alongside the dehumanization of Latin American women represented as highly sexualized and the disparagement of nonbinary users by binning them into a cluser of being `women-lite.' 

Our findings extend prior research into the harms caused by T2Is. Our work most closely builds upon \citet{ghosh2023person}, as we demonstrate the real-world epistemic experiences of users of GAI tools facing the stereotypes they observe (such as sexualization of Latin American women), while also grounding these in a framework of harm and providing additional novel experiences. This is an important extension, especially when considering how this work can provide testimonial evidence in contributing towards policy formation informing T2I design, by demonstrating how real people face T2I harms. Furthermore, while we do not study harms towards a single community in depth like \citet{qadri2023ai} and also document a similar pattern of South Asian communities being depicted as impoverished, we extend this work by demonstrating how this depiction persists across a much larger number of images than they were working with. Our research extends the state-of-the art by showing how Stable Diffusion, a globally popular model being used by millions of users daily \cite{stablediffusionusage}, causes so widespread representational harms -- namely stereotyping, erasure, disparagement, and dehumanization \cite{dev2021measures}. While our sample size of 14 interviewees and 133 crowdsourced participants might not be enough to claim generalizability, we determine alarming trends across a diverse range of genders and nationalities. 

Although Stable Diffusion was beta-tested with over 10,000 users across 1.7 million images \cite{stablediffusion_history}, our work reveals that there is still a significant gap between user expectations and model performance. This is concerning for a service with such global popularity and indicative of subpar user-testing before launching into production or, in the most generous scenario, not continuing periodic iterative testing after launch. This disconnect between user expectations and outputs also has significant implications for Stable Diffusion and other T2Is being embedded into downstream tasks. Consider T2I usage in the generation of videos targeted towards a particular demographic: if Stable Diffusion produces images which videomakers think their target audience would relate with but the videos contain the aforementioned disconnect, the content will be poorly received. While Stable Diffusion represents a cheap and easy-to-use tool for tasks such as content creation, users must be cognizant of the disconnect between its outputs and their expectations, as the success of the task might heavily depend on how strong this disconnect is with the target audience. 

This concern elevates when the sentiment moves from users being generally dissatisfied with outputs to feeling actively harmed, as such harms can have short and long-term effects on an individual's physical and mental health. For example, when Stable Diffusion generates images of individuals of nonbinary gender as `women-lite' or when other T2Is depict them as having lilac/purple quiffs \cite{wired}, users who might not be fully aware of the fallibility of such services might be adversely affected. As \citet{404media} writes, ``for AI to depict stereotypical images of what it means to ‘look trans/nonbinary’ has the potential to cause real harms upon real people ... Especially for young people, who might be seeing such images more and more in their daily media diets, this can create an unhealthy impression that there is a ‘correct’ way to present oneself as trans/nonbinary.” The potential for Stable Diffusion and other T2Is causing widespread harms is something to be taken seriously, and addressed through alternative design approaches. 

\section{A Harm-Aware Approach to T2I Design}

Based on our findings, we advocate for a \textit{harm-aware} approach towards the design of GAI tools such as Stable Diffusion. We propose this approach as complementary to other calls for community-centered research in GAI design \cite[e.g.,][]{ghosh2023chatgpt, gadiraju2023wouldn, qadri2023ai}, but extend these by focusing on harms caused. 

\subsection{Centering Harm Reduction, instead of Retroactive Efforts}

Common approaches towards designing T2Is, among other systems, typically operate through identifying a problem or a set of pain points shared across a set of potential users, and consider design solution towards rectifying those pain points. For T2Is, such approaches might also involve patterns of user-testing where users might be shown outputs for prompts and asked to evaluate the `accuracy'. Products and GAI tools designed over such approaches typically lead to the exposition of potential shortcomings, such as inaccessibility and harms caused, after their deployment. Indeed, this paper is one such example, as we explicate representational harms within Stable Diffusion. Borrowing on approaches within accessibility research that argue for systems and products being designed to be accessible rather than adding in accessibility features after-the-fact \cite{mack2022anticipate}, we propose a harm-aware approach centering and prioritizing the mitigation of harms during the design process, instead of mitigating emergent harms after deployment. 

One of the principles of a harm-aware approach would be to not only prioritize extensive user involvement all through the design process with several rounds of user testing and iteration, but also recognize that people are more than just `users' who simply provide data. Rather, they are `humans' with extensive lived experiences and individual and collective values \cite{gasson2003human} that are inextricable from their inputs. A harm-aware approach would prioritize design decisions that honor those lived experiences and values \cite{chancellor2023toward}, with the central principle of not bringing harm to those experiences and values through designed systems. 

A design team following such an approach must also employ within their ranks one or more ethicists. One of the primary criticisms of fields that focus on fairness in designed systems is that they are highly technical \cite{laufer2022four}, and often lose touch with the underlying ethical principles they are implicitly working towards by considering that because they know the principles of ethical design, they should be fine \cite{mittelstadt2019principles}. With ethicists collaborating with engineers and product designers, a harm-aware design team would address the problem of harms being \textit{sociotechnical} \cite{cooper1971sociotechnical}, and thus evaluating the overall performance of the system based on both the technical and social aspects \cite{bowker1997social} such that failing to consider either component would lead to an inaccurate assessment and miss opportunities for design improvement \cite{badham_wall_2006}. Ethicists could also contribute towards building a strong understanding within the team of the societal structures and hierarchies of privilege that would govern T2I design and exist in the world in which the T2I would be deployed, as well as be well positioned to `examine power' \cite{d2020data} within the design process by taking stock of which identities are represented and which are not. In sum, ethicists working with designers and engineers would be the best placed to identify potential types of harms that a T2I can cause, anticipate the sources of such harms, and advise on what the most ethical approach towards harm reduction could be \cite{birhane2021algorithmic}. 

\subsection{Harm-aware Data Practices}

Although harms caused by T2I outputs are rooted in much more than faulty data and the conception of this problem as `bias in, bias out' is far too simplistic \cite{d2020data, birhane2021algorithmic}, data plays a not-so-insignificant part in producing such harmful outputs. Drawing from community-centered approaches \cite[e.g.,][]{gadiraju2023wouldn, mack2024they, qadri2023ai}, we outline principles for harm-aware data practices. Such practices would heavily involve community-centered approaches to data collection \cite{ghosh2023chatgpt, qadri2023ai}, providing agency to individuals in defining representations of their identities within GAI tools. Such data collection should also keep in mind potential harms that can be caused when resultant datasets are used to train GAI tools, and would require the design team to work closely with communities in anticipating such harms. A key step is understanding how data gathered from different communities intersect, and which representations are more prominent \cite{bender2018data}. Designers should collect volumes of data proportionate with community sizes, but also be careful to not harm smaller communities with minimal representation in datasets. 

It is also important to work with multiple groups within communities across different sessions, since not everyone will have the same perspective on identifying with the community. For instance, \citet{qadri2023ai}'s work on South Asian representation in T2Is found that while Pakistani and Bangladeshi participants surfaced the stereotype of `Indian-ness' within depictions of South Asian people, Indian participants did not. This paper also avoids asking participants for generalized claims from their participants to speak on behalf of their cultures and perform the same homogenization we criticize Stable Diffusion for performing. This is also true for small and hyperlocal groups, since even those individuals might have different ideas of group membership and what constitutes a harmful depiction of their community. 

It is also important to consider that for some communities, collecting data from them in any capacity is one of the ways in which harm can befall them. For instance, there is extensive evidence that many transgender individuals do not feel safe in marking that as their gender in official reports or sharing it widely with their families, for fear of being at the receiving end of hate crimes and legal ramifications \cite{hrw_trans}. Data collection is often the province of the powerful, and a truly harm-aware approach would recognize and afford agency to users or communities who do not wish to participate in the T2I design process \cite{d2020data}. When consent to collect data is provided, it must be conducted in a considerate manner, ensuring that contacted individuals have adequate opportunities to provide and withdraw consent from participation \cite{d2020data, ghosh2023chatgpt}. A harm-aware approach will also keep in mind that simply collecting data from communities is a form of epistemic extractivism \cite{grosfoguel2019epistemic}, and should not be done without meaningfully (as defined by the community themselves) contributing back to the community \cite{ghosh2024misgendering}, especially in the case of multiply marginalized or highly vulnerable communities.

\subsection{Harms Caused as an Evaluation Metric}

Furthermore, a harm-aware approach would also consider harms caused, or potential for harms being caused, as an important metric of evaluating GAI tools under design. Currently, GAI tools like Stable Diffusion are evaluated through a combination of qualitative and quantitative techniques. Qualitative evaluation procedures include human assessments of image composition and alignment with prompts, as well as spatial relations \cite{salaberria2024improving}. These are balanced with quantitative metrics such as Frechet Inception Distance (FID) to measure the similarity between datasets of images \cite{borji2022generated} and CLIP score to measure compatibility of text-image pairs. A harm-aware approach would place harms caused at the same importance as these metrics, and dictate that a developed GAI system should score sufficiently low on metrics of harm before deployment.

It is thus important to talk about defining the metric of harms caused. Borrowing from \citet{blodgett2020language}, we urge developers of harm-aware GAI tools to define harms in their specific contexts keeping in mind which harms to consider, what ways such harms might happen within their systems, and who might suffer the brunt of such harms more than others. To determine harms to consider, it is important to remember that while most studies (including this one) focus on representational harms such as stereotyping and erasure -- e.g., \citet{qadri2023ai} focuses on the stereotyping and erasure of South Asian identities, \citet{mack2024they} focuses on these harms in the context of disability, etc. -- there also exist the slate of allocational harms \cite{barocas2017fairness} which will most likely transpire when GAI tools are embedded into downstream services, and there remains the potential that novel types of harm might emerge through future research. When considering which users might experience harms caused by GAI tools being designed, it is important to consider the context in which the tool is being deployed and prioritize harms accordingly, i.e., while a GAI system that is being deployed globally should prioritize mitigating harms along the lines of identities such as race, gender, and disability status which transcend national borders, a system which might be designed in a more local context should prioritize identities that might be location specific, e.g., a GAI tool being designed for use exclusively in India should consider, alongside the aforementioned aspects of identity, how to mitigate casteist harms which are prevalent in Indian contexts. 

Such an approach must also define how the metric around harms caused will be measured. \citet{dev2021measures} proposes a few questions to consider when defining such a metric, such as will the metric be absolute or relative, and can the metric imply an absolute absence of harm within the GAI tool. We advocate for the use of a relative metric of measuring harm that recognizes that individuals and groups with one or more historically marginalized identities are more likely to encounter harms, which necessitates a stronger weight being provided to harms they encounter as opposed to potential harms faced by historically privileged identities. This is \textit{not} to say that designers should be building implicit hierarchies of marginalized identities, e.g., somehow comparing the impact of racist harms against sexist harms, etc., but rather that they should assign higher value to designing for historically marginalized populations, in opposition to traditional practices of centering the voices of traditionally privileged groups. Furthermore, we currently do not believe it to be possible to achieve an absolute absence of harm within the outputs of GAI tools. To be human is to have biases \cite{miceli2022studying} and despite an individual's best efforts to keep their biases at bay when working on something, it is impossible for them to guaranteedly say that none of their biases, both conscious and unconscious, influenced their work. Since the accumulation of biases contribute to the production of harmful outputs, it is difficult to reliably claim an absolute absence of the potential for a designed GAI tool to cause harm. We believe that designers following a harm-aware approach carefully work to minimize the types of harms caused and the number of communities they affect, and upon deployment, acknowledge the possibility of some harms still coming through in outputs by warning users and preparing policies to document harms towards which reduction efforts can focus. 

\subsection{Iterative Development of Harm-Aware GAI Tools, and Considering Downstream Tasks}

A harm-aware approach to the design of GAI tools does not stop at the deployment stage. Rather, such an approach dictates that designers and product managers rigorously monitor the usage of their GAI tool and proactively solicit user perspectives on performance and harms caused. While a GAI tool might be put through the most rigorous of tests during the design phase, the fruits of such rigor must concede to real-world evaluations by users outside the individuals involved within the design process. Designers must invite and continually monitor user experiences, and be prepared to iterate in the face of an unanticipated harm occurring. This should include a willingness for deployed tool which is actively causing harm to be entirely taken down, as was the case with Microsoft's WizardLM 2 which was deployed and then removed because users flagged its highly toxic outputs \cite{404msft}, or have its capabilities reduced, e.g., Google's temporary suspension of their T2I generating human faces after users found it to generate historically inaccurate images \cite{gautam2024melting}.

Additionally, designers who adopt a harm-aware approach to GAI tools should also consider the potential set of downstream tasks for which their tools can be used. If a GAI tool or model is meant to be deployed for global use, such as GPT-3+ or Stable Diffusion, then designers should not only take extra care in ensuring that the potential harms caused by their outputs be as minimal as possible, but also consider that they cannot possibly account for every possible usecase that a third-party organization or individual might have for their tool. However, for tools with stronger licensing practices that require third-party organizations to acquire the rights to use their tools in downstream tasks, GAI designers must thoroughly vet the proposed applications of their tool in such tasks. As experts in their own tool, designers must carefully evaluate the prospect of harms occurring in downstream tasks and consider that a central factor in deciding whether their tool should be embedded in such tasks. 

Similar to \citet{ghosh2023chatgpt}, we invite practitioners and designers of current/future GAI tools to \textit{try} out this harm-aware approach towards GAI design. We are actively working on refining this further and do not claim this design process to be foolproof. We invite organizations with sufficient resources and interest in designing GAI tools to try this approach with their own adaptations, and we are willing to work with them on this. In the spirit of iterative design, we believe that this approach will only improve with iteration. 

\section{Limitations and Conclusion}\label{sec:limitations}

A limitation of our work is around interview sampling, because our participants skew female. Our female participants not seeing themselves represented in outputs to the `Person from X Country' prompts which showed mostly male-presenting faces is thus a direct result of this sampling. While we do not devalue the finding because it highlights the masculine default within Stable Diffusion \cite{ghosh2023person}, we find this pertinent to note. 

In this paper, we explore user perspectives on images of human faces generated by the popular T2I Stable Diffusion. We uncover a deep disconnect between user expectations of what a `Person' would look like and what Stable Diffusion depicts for this prompt, highlighting the need for more extensive user studies. We also document Stable Diffusion causing the algorithmic representational harms of stereotyping, disparagement, dehumanization, and erasure \cite{dev2021measures}, through first-hand comments from direct users of Stable Diffusion. We conclude with a proposal for a harm-aware approach to designing Text-to-Image Generators.
\newpage

\section*{Ethical Considerations, and \\Adverse Impact Statement}

One of the ethical considerations we wish to note is the potential adverse impact of the images shown in this paper being made available on the Internet and associated with the prompts mentioned in the captions. For instance, it is problematic to upload Figure \ref{fig:venezuela} associated with the phrase `Woman from Venezuela', because it can contribute to spreading the same stereotype we are calling out in this paper. Furthermore, if this text-image pair is incorporated into the training dataset of a GAI tool, then the outputs of that tool may also be more likely to spread the same harm of stereotyping we document within Stable Diffusion. Therefore, pursuant to \citet{ghosh2023person}, we will only share blurred versions of these images when the paper is being uploaded on the Internet.

\section*{Researcher Positionality Statement}

The first author is a brown cisgender international student to the United States, hailing from India. The second author is a white cisgender American woman, who is a part of the Hispanic Latinx diaspora. The third author identifies as an American of Bulgarian-Turkish origin, having spent half of their life in the United States.

\section*{Acknowledgements}

This work was supported by the U.S. National Institute of Standards and Technology (NIST) Grant 60NANB23D194. Any opinions, findings, and conclusions or recommendations expressed in this material are those of the authors and do not necessarily reflect those of NIST.

We are also appreciative of our reviewers for their thoughtful feedback, which undoubtedly made this paper richer than it otherwise would have been. 

\bibliography{aaai24}
\begin{appendices}
\renewcommand{\thesection}{\Alph{section}}
\section{: Data}\label{app:data}

\begin{figure}[h]
  \centering
  \fbox{\includegraphics[width=\linewidth]{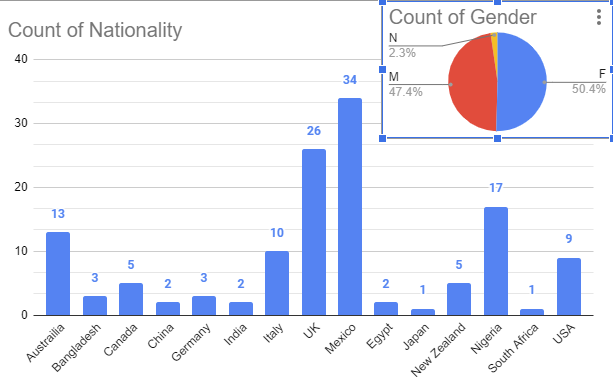}}
  \caption{Demographic data from Prolific participants, containing information on self-reported gender and nationalities across 133 respondents.}
    \label{fig:prolific}
\end{figure}

From Prolific, we collected a total of 147 responses, but deemed only 133 to be valid because the remaining 14 did not complete their respective Diamond ranking tasks, i.e. ended the study without filling up the 20 cells in the Diamond. We also present demographic information across the 133 responses. In terms of region, we received 48 responses from North America [5 from Canada, 9 from the US, and 34 from Mexico], 39 from Europe [3 from Germany, 10 from Italy, and 26 from the UK], 20 from Africa [1 from South Africa, 2 from Egypt, and 17 from Nigeria], 18 from Oceania [5 from New Zealand, and 13 from Australia], and 8 from Asia [1 from Japan, 2 from China, 2 from India, and 3 from Bangladesh]. In terms of self-identified gender, we received 67 responses from female respondents, 63 from male respondents, and 3 from those who identified as nonbinary gender. This information is presented in Figure \ref{fig:prolific}. 

Furthermore, we also provide in Table \ref{tab:codes} information about the most salient codes and themes that emerged during our thematic analysis process, as well as researcher-determined definitions of those.  

\begin{table*}[h]
  
  \begin{tabular}{p{5.3cm}p{10.5cm}}
    \toprule
    \textbf{Code/Theme} & \textbf{Definition} \\
    \midrule
    Self-reported gender & Participant's self-reported gender information. \\
    Self-reported nationality & Participant's self-reported nationality information. \\
    Personal Description & Examples where a participant self-describes how they look or appear, as if they were writing a caption for themselves. \\
    GAI Strengths & Perceived strengths or positives of GAI tools mentioned by participants, including examples of personal experiences. \\
    GAI Weaknesses & Perceived drawbacks or negatives of GAI tools mentioned by participants, including examples of personal experiences. \\
    Initial Expectations & Participants' expectations about the quality of GAI outputs to study prompts \textit{before} starting Diamond Ranking task. \\
    Self-Representation & Participants' experience of feeling represented within displayed output. \\
    Self-Erasure & Participants' experience of not feeling represented within displayed output. \\
    Output Satisfaction & Participant mentions feeling satisfied (or displays satisfaction with nonverbal cues) with displayed output. \\
    Output Dissatisfaction & Participant feeling dissatisfied (or displays dissatisfaction with nonverbal cues) with displayed output. \\
    Output Anger & Participant feeling angry (or displays anger with nonverbal cues) with displayed output.\\
    Sexualization & Participant notices evidence of sexualization within displayed output. \\
    Enby as `Women-lite' & Participant notices evidence of gendernonbinary individuals being displayed as derivative of female-presenting faces. \\
    Diversity & Participants comments on the diversity (or lack thereof) of people and appearances within displayed output.\\
    Repetition & Participants notice features or aspects repeated across multiple images within displayed output. \\
    
  \bottomrule
\end{tabular}

\caption{Salient codes/themes found within interview transcripts, as well as researcher-determined definitions.}
\label{tab:codes}
\end{table*}
Finally, all images used in this study can be found here: https://drive.google.com/drive/folders/18ggMPMn\_is6n\_5sJ0xLFqsnWEwcvWClY?usp=sharing. 
\end{appendices}
\end{document}